\documentclass[widetext,showpacs,preprintnumbers,amsmath,amssymb]{revtex4}

\usepackage{graphicx}
\usepackage{dcolumn}
\usepackage{bm}

\begin{document}
\title{Multi-directed Eulerian growing   networks}
\author{A. P. Masucci}
\author{G. J. Rodgers}
\affiliation{%
Department of Mathematical Sciences, Brunel University, Uxbridge,
Middlesex, UB8 3PH, United Kingdom}%

\date{\today}
\begin{abstract}
We introduce and analyze a model of a multi-directed Eulerian
network, that is a directed and weighted network where a path exists
that passes through all the edges of the network once and only once.
Networks of this type can be used to describe information networks
such as human language or DNA chains. We are able to calculate the
strength and degree distribution in this network and find that they
both exhibit a power law with an exponent  between 2 and 3. We then
analyze the behavior of the accelerated version of the model and
find that the strength distribution has a double slope power law
behavior. Finally we introduce  a non-Eulerian version of the model
and find that the statistical topological properties remain
unchanged. Our analytical results are compared with numerical
simulations.

\end{abstract}
 \maketitle

\section{\label{sec:level1}Introduction.\protect}
Many naturally occurring systems appear as  chains of repeated
elements. Such systems, such as human language, DNA chains, etc..,
often encode and transport information. Markov processes have been
adopted to model those chains\cite{M}. Unfortunately Markov chains
are not able to describe long range correlations that exist within
these structures. Thus complex growing networks appear to be a more
suitable modeling tool.

In this paper we study written human language as a complex growing
network. Since the discovery by Zipf \cite{1} that language exhibits
a complex behavior, and the application of Simon's theories\cite{3}
to growing networks\cite{4}, this topic has been examined by a
number of scientists \cite{l1,l2,l3,l4}.

A useful  way to build a network from a text is to associate a
vertex to each sign of the text, that is both words and punctuation,
and to put a link between two vertices if they are adjacent in the
text. In a previous paper \cite{ap} we showed that it is necessary
to consider a directed and weighted network to understand the
topological properties of this language network, in which the weight
of each link in the network represents the number of directed links
connecting two vertices. Directed links in such a network are
necessary since they need to describe systems in which a syntax is
defined and where attachments rules between the objects are not
reflexives \cite{ap}.

When networks are built in this way, from a chain of repeated
elements,  a weighted adjacency matrix is obtained that is  well
known graph in graph theory: the \emph{multi-Eulerian
graph}\cite{g}. Eulerian means that there exists a path in the graph
passing through all the links of the  network once and only once,
while the prefix:"\emph{multi}" refers to the fact the adjacency
matrix allows multiple links between two vertices.

 In order to describe the evolution of a multi-directed graph  we need to introduce the formalism of weighted
networks\cite{w,w1}. These are characterized by a weighted adjacency
matrix $W=\{w_{ij}\}$ whose elements $w_{ij}$ represent the number
of directed links connecting vertex $i$ to vertex $j$. We define the
degree $k_i^{out/in}$ of vertex $i$ as the number of out/in-nearest
neighbours of vertex $i$ and we have
$k_i^{out/in}=\sum_j\Theta(w_{ij/ji}-\frac{1}{2})$. We define the
out/in-strength $s_i^{out/in}$ of vertex $i$ as the number of
outgoing/incoming links of vertex $i$, that is
$s_i^{out/in}=\sum_jw_{ij/ji}$. Analytically the Eulerian condition
means that the graph must be connected and it must have
$s_i^{in}=s_i^{out}$ for every $i$.

In this work we first  develop and analyze  a model for a general
multi-directed Eulerian growing  network. Then, since human language
is an accelerated growing network,we extend our model to its
accelerated version, and find results similar to those in \cite{l1}.
More recent works on accelerated growing networks can be find in
\cite{p,x}. To conclude we introduce and analyze the non-Eulerian
version of our model. This last step allows us to build a directed
network without initial vertex attractiveness. As far as we are
aware, this is the first time a model for directed networks has been
proposed without the help of this ingredient. The resulting power
laws exponents, tunable between $2$ and $3$, are very interesting
since they fit with those found within most of the real
networks\cite{4,d}.

\section{Model A}

\begin{figure*}
\includegraphics[width=4cm]{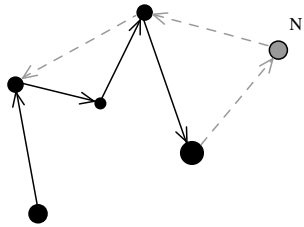}
\caption{\label{f1} Growth mechanism for  model A with $m=1$. Dashed
grey arrows represent $m+2$ newly introduced edges. }
\end{figure*}

First we introduce a model for the multi-directed Eulerian growing
network which we will call  $Model A$. The Eulerian condition
(hereafter EC) states that every newly introduced edge has to join
the last connected vertex, so that every newly introduced in-link
implies a constrained out-link from the last connected vertex. This
is equivalent to say that
$s_i^{in}=\sum_jw_{ij}=\sum_jw_{ji}=s_i^{out}=s$, for every $i$,
with the global constraint the network must be connected
(Fig.\ref{f1}). With the last condition  our calculations become
easier since we have to consider one quantity, that is $s$, instead
of two.

We start with a chain of $2m$  connected vertices. At each time step
we create a new vertex and $m+2$ new directed edges (Fig.\ref{f1}).
At each time step

a- The new vertex will acquire one in-link with the constraint that
the network must respect the EC.

b- The remaining $m+1$ in-links will be attached to old vertices
with probability proportional  to their in-strength with the
constraint that the network must respect the EC.

To calculate the strength distribution for the model, we use the
fact that with  the EC the in-strength will be exactly the same as
the out-strength distribution. We write the equation for the
strength evolution $s(t,t_i)$ at time $t$ for the vertex born at
time $t_i$ as:

\begin{equation}\label{1}
\frac{ds(t,t_i)}{dt}=(m+1)\frac{s(t,t_i)}{\sum_i s(t,t_i)}.
\end{equation}

 The right hand side of the last equation takes into account that $m+1$ vertices
 acquire a link
  with probability proportional to their normalized strength $\frac{s_i}{\sum
  s_i}$. Considering that the total number of in/out-links at time
  $t$ is $\sum_i  s(t,t_i)=(m+2)t$ and integrating Eq.\ref{1} with the initial condition  $s(t_i,t_i)=1$ we obtain
\begin{equation}\label{2}
s(t,t_i)=\left(\frac{t}{t_i}\right)^{\frac{m+1}{m+2}}.
\end{equation}

Using the fact that

\begin{equation}\label{ps}
 P(s,t)=-\frac{1}{t}\frac{\partial
t_i}{\partial s(t,t_i)}|_{s(t,t_i)=s(t)}
\end{equation}

  from Eq.\ref{2} we obtain:
\begin{equation}\label{3}
P(s,t)=\frac{m+2}{m+1} s^{-\frac{2m+3}{m+1}}
\end{equation}
which is a stationary power-law distribution with exponent between
 $2$ and $3$. In particular it will be $3$ for $m=0$, and it will
 tend to $2$ for increasing values of $m$.

In order to calculate the degree distribution we consider that each
time the strength of a vertex increases by 1, the degree of the
vertex increases if and only if the vertex links with a new
neighbor. This process implies higher order correlations. We will
approximate this process as an uncorrelated one and compare our
results with simulations. Hence the equation governing the evolution
of the degree is

\begin{equation}\label{4}
\frac{dk(t,t_i)}{dt}=\left[1+m\left(1-\frac{k(t,t_i)}{t}\right)\right]\frac{s(t,t_i)}{\sum
s(t,t_i)}.
\end{equation}

To understand this equation we have to notice that the degree of a
vertex grows at a rate proportional to its normalized strength, as
in Eq.\ref{2}, but, when the strength of a vertex increases by 1,
the probability that the degree of the vertex $i$ increases by 1 is
$(1-k(t,t_i)/t)$. In fact $k(t,t_i)$ is the number of nearest
neighbors of vertex $i$, while $t$ represents the total number of
vertices at time $t$. Note that for $m=0$, $k(t,t_i)=s(t,t_i)$ as we
would expect.

We substitute Eq.\ref{2} in  Eq.\ref{4} and we integrate it to
obtain

  \begin{equation}\label{5}
  k(t,t_i)=\frac{(m+1)m^{m+1}}{t_i^{m+1}}\exp{\left(\frac{mt^{-\frac{1}{m+2}}}{t_i^{\frac{m+1}{m+2}}
  }\right)}\left\{\Gamma\left[-(m+1),\frac{m}{t_i^{\frac{m+1}{m+2}}
  }t^{-\frac{1}{m+2}}\right]+C\right\}
  \end{equation}

    where $\Gamma(a,b)$ is the incomplete Gamma function and $C$ is an integration constant to be determined by
    the initial conditions $k(t_i,t_i)=1$. For $m=0$ the right hand side of Eq.\ref{5} is an indefinite form.
    Nevertheless, taking the limit for $m\rightarrow 0$, we find
    again the result of Eq.\ref{2}, as we predicted.

    We are interested in the behavior of the network for large values of $t$, so that we expand
    the first incomplete Gamma function for small values of its second
    argument. Then  we take the limit of the expression for $t\rightarrow \infty$ and
    obtain

  \begin{eqnarray}\label{6}
  k(t,t_i)\approx   \left(\frac{t}{t_i}\right)^{\frac{m+1}{m+2}}.
  \end{eqnarray}

  Using again Eq.\ref{ps} for the degree we get

  \begin{equation}\label{7}
  P(k,t)\propto k^{-\frac{2m+3}{m+1}}
  \end{equation}
which is again a stationary power-law distribution with exponent
 between 2 and 3.

 \begin{figure}[!ht]\center
         \includegraphics[width=0.49\textwidth]{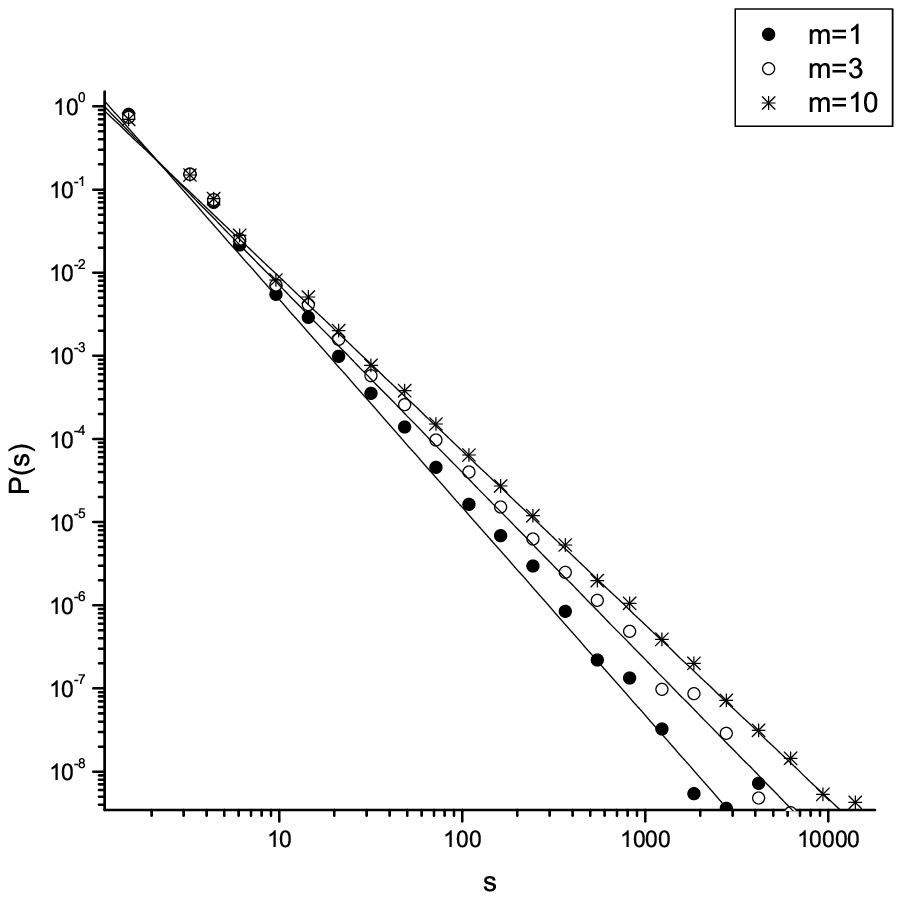}
         \includegraphics[width=0.49\textwidth]{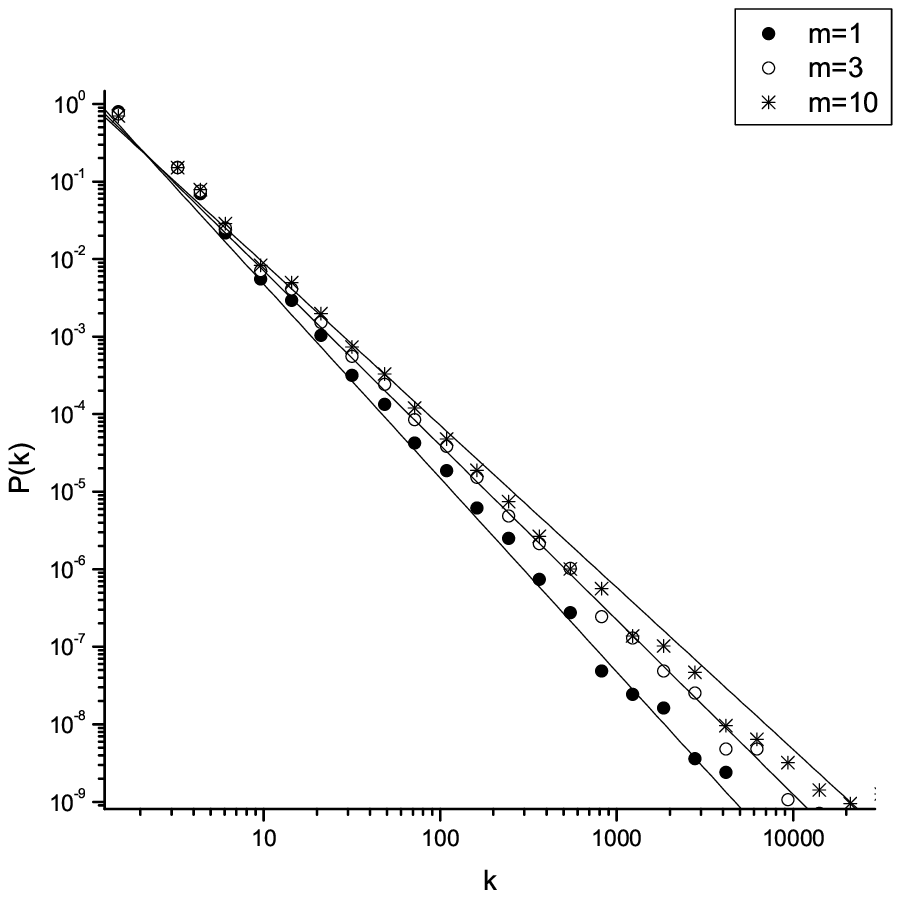}
         \caption{\label{f2} Results from a 250000 vertices simulation
         for  model A for different values of $m$. On the left
         the strength distribution is compared with Eq.\protect\ref{3}. On the
         right the degree distribution is compared with Eq.\protect\ref{7}.
         Simulation results are points while lines represent the
         analytical results.
         }
 \end{figure}

\begin{figure*}
\includegraphics[width=9cm]{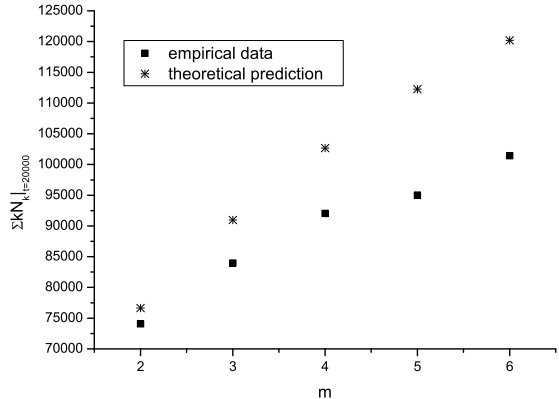}
\caption{\label{f3} Comparison between the numerical integration of
Eq.\protect\ref{5} and numerical simulations. This integral
represents the number of occupied cells of the adjacency matrix
evaluated for $t=20000$ and varying values of $m$, while the
empirical data count the effective number of occupied cells.}
\end{figure*}

To check Eq.\ref{5} we  integrated it for different values of $m$
and fixed $t$. This integral represents the number of occupied cells
of the adjacency matrix and can be compared with results obtained by
simulations. The results are shown in Fig.\ref{f3}. As we can see
the uncorrelated approximation is very good for small values of $m$,
but it fails to reproduce the behavior of the system for larger
values of $m$, when correlations are stronger.

In Fig.\ref{f2} we plot the simulations results against Eq.\ref{3}
and Eq.\ref{7} for different values of $m$. In the case of the
strength distribution the goodness of the fit is excellent, while in
the case of the degree distribution the approximate result of
Eq.\ref{7} gives just an approximate fit for large values of $m$,
and it is because of the growing strength of correlations in the
network for large values of $m$.

\section{Model B }

In this section we build and analyze a multi-directed accelerated
growing Eulerian network that is  an accelerated version of the
previous model and we will call it $Model B$. In order to do this we
replace the constant addition of $m$ edges at each time step with a
number of edges $m'$ that grows linearly with time, that is
$m'=\alpha t$. In this way at every time step we have an increasing
number of edges added to the network.  The obtained results and the
used techniques are similar to the ones used in \cite{l1}.
Nevertheless this extension of the previous model is designed to get
closer to the topology of real language networks, as they display an
accelerated evolution, and it is important for completeness in the
discussion of the subject.

Keeping this in mind we can describe our modified model. We start
with a chain of some  connected vertices. At each time step we
create a new vertex and $\alpha t+2$ new directed edges
(Fig.\ref{f1}). In particular at each time step

a- The new vertex will acquire one in-link with the constraint the
network must follow EC.

b- The remaining $\alpha t+1$ in-links will be attached to old
vertices with a probability proportional  to their in-strength with
the constraint the network must follow EC.

The coefficient $\alpha$ will be chosen to fit with that found in
real language networks\cite{ap}.

The equation for the strength evolution of the strength of vertex
$i$ is
\begin{equation}\label{8}
\frac{ds(t,t_i)}{dt}=(\alpha
t+1)\frac{s(t,t_i))}{\int_{0}^tdt_is(t,t_i)}.
\end{equation}
The right hand side of the last equation takes into account that
$\alpha t+1$ vertices can
 acquire a link
  with probability proportional to their normalized strength $\frac{s(t,t_i)}{\int_{0}^tdt_is(t,t_i)}$.
  The integral at the denominator  in the right hand side of Eq.\ref{8}
represents the total strength of the network and is
$\frac{1}{2}\alpha t^2+2t$.

Solving Eq.\ref{8} with initial condition $s(t_i,t_i)=1$ we obtain

\begin{equation}\label{9}
s(t,t_i)=\left(\frac{t}{t_i}\right)^\frac{1}{2}\left(\frac{\alpha t
+4}{\alpha t_i+4}\right)^\frac{3}{2}.
\end{equation}

To calculate the strength distribution we use the fact that
\begin{equation}\label{PS}
P(s,t)=-\frac{1}{t}(\frac{\partial s(t,t_i)}{\partial
t_i})^{-1}|_{t_i=s(t,t_i)}
\end{equation}
 and we get

\begin{equation}\label{10}
P(s,t)=\frac{2t_i}{ts}
\end{equation}
where $t_i(s,t)$ is the solution of Eq.\ref{9}.

This distribution has two regimes separated by a cross-over given
approximatively by $s_{cross}\approx (\alpha
t)^\frac{1}{2}(\frac{\alpha }{8}t+\frac{1}{2})^\frac{3}{2}$.

Below this point Eq.\ref{10} scales with a power law as
\begin{equation}\label{11}
P(s)\propto s^{-\frac{3}{2}}
\end{equation}

while, for $s>s_{cross}$,
\begin{equation}\label{12}
P(s)\propto s^{-3}.
\end{equation}

\begin{figure*}
\includegraphics[width=9cm]{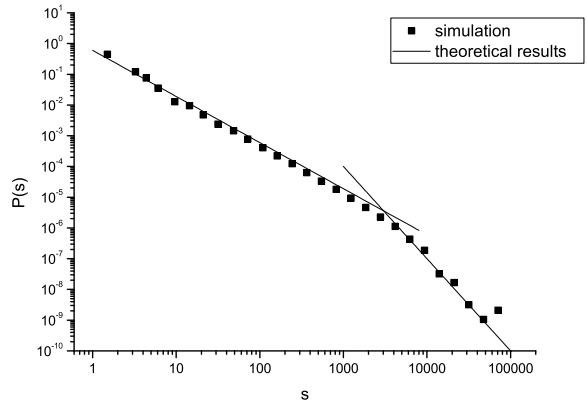}
\caption{\label{f4} Comparison between the numerical simulation for
model B in a  network of 50000 vertices and Eq.\protect\ref{11} and
Eq.\protect\ref{12}.}
\end{figure*}

These results are well confirmed by numerical simulations as shown
in Fig.\ref{f4}.

\section{Model C}

To complete this work we introduce a non-Eulerian version of model A
and we call it $Model C$.

We start with $2m$ randomly connected vertices. At each time step we
create a new vertex and $m+2$ new directed edges. In particular at
each time step

a- The new vertex will acquire one in-link and one out-link.

b- The remaining $m+1$ out-links will be attached to old vertices
with probability proportional to their out-strength.

c- The remaining $m+1$ in-links will be attached to old vertices
with probability proportional to their out-strength.

For this model the same equations apply as with the Eulerian Model A
with the same arguments, so that it displays equivalent topological
properties, that is weight, strength and degree distributions. The
main difference at this level of observation is that in the Eulerian
case $s^{in}=s^{out}$ in an exact sense, while in this case this
condition holds only on average.

\section{Conclusions}

In this work we contextualize phenomena that manifest as a
continuous chain of repeated elements in a novel way, within the
framework of network theory. We show that such phenomena, such  as
human language, DNA chains, etc.., are  described by Eulerian
graphs. Eulerian graph topology ensures that every newly connected
vertex of the network is connected to the last linked vertex. So we
introduce and analyze different kinds of growing networks built to
produce an Eulerian graph. We are able to find the main topological
properties for this kind of network and we find that the resulting
exponents for the strength and degree distributions are compatible
with those of real networks. We then extend our model to a
non-Eulerian one.

It is worth noting that, in the context of the standard network
analysis, no striking differences emerge between the Eulerian
network and its non-Eulerian counterpart. We  performed a clustering
coefficient analysis, but it was not worth showing it, since the
differences between the average number of triangles formed in the
network in the Eulerian and non-Eulerian case didn't differ
significatively. Even a Shannon entropy analysis wouldn't define any
relevant difference between the two different growing mechanisms,
since it is based on the frequency of the elements more than on
their structural organization. Considering this and the fact that
the geometry of the two different networks are so dissimilar,
dissimilar as a tree and a chain, we would like to emphasize the
lack of statistical tools, in network theory, to characterize in a
significant statistical way the different morphologies of different
networks.

This work is mainly focused on the analysis of written human
language, but it is also important for the study of directed and
weighted growing networks. An important extension of these models,
that could be taken into consideration for further investigations,
is the growth of a network governed by local growing rules. We
showed in a previous work\cite{ap} that local growing rules are
important to reproduce interesting features of human language and
must be taken into account to generate a syntax-like structure.

\begin{acknowledgments}
This research is part of the NET-ACE project (contract number 6724),
supported by the EC.

\end{acknowledgments}


\thebibliography{apsrev}

\bibitem{4} A.L. Barabasi, R. Albert,  H. Jeong, Physica A \textbf{272}, 173 (1999).
\bibitem{w} A. Barrat, M. Barthelemy, R. Pastor-Satorras, A.
Vespignani, Proc. Natl. Acad. Sci. USA \textbf{101}, 3747 (2004).
\bibitem{w1} A. Barrat, M. Barthelemy,  A. Vespignani, Phys. Rev. E \textbf{70}, 066149 (2004).
\bibitem{g} G. Chartrand, L. Lesniak, \textit{Graphs \& digraphs}, Chapman \&
Hall, 1996.
\bibitem{d} S.N. Dorogovtsev, J.F.F. Mendes, A.N. Samukhin, Phys. Rev. Lett. \textbf{85}, 4633
(2000).
\bibitem{l1} S.N. Dorogovtsev, J.F.F. Mendes, Proc. Roy. Soc. London B \textbf{268}, 2603 (2001).

\bibitem{l2} R. Ferrer i Cancho, R.V. Sole,  Proc. Roy. Soc. London B \textbf{268}, 2261 (2001).

\bibitem{ap} A.P. Masucci, G.J. Rodgers, Phys. Rev. E \textbf{74}, 026102 (2006).

\bibitem{l3} M.A. Montemurro, P.A. Pury,  Fractals \textbf{10}, 451 (2002).
\bibitem{p} P. Sen, Phys. A, \textbf{346}, 139 (2005).

\bibitem{3}H.A. Simon, Biometrika \textbf{42}, 425 (1955).
\bibitem{M} O.V. Usatenko, V.A. Yampol'skii,  Phys. Rev. Lett. \textbf{90}, 110601 (2003).

\bibitem{1}{G.K. Zipf, \textit{Human Behaviour and the Principle of Least Effort}, Addison-Wesley Press, 1949.}
\bibitem{x} X. Yu, Z. Li, D. Zhang, F. Liang, X. Wang, X. Wu,  J. Phys. A \textbf{39}, 14343
(2006).
\bibitem{l4} There is a large on-line bibliography on linguistic and cognitive networks at \textit{http://complex.ffn.ub.es/~ramon/linguistic\_and\_cognitive\_networks.html\ }
\end{document}